\documentclass{PoS}
\usepackage{lineno}
\usepackage{amsmath}

\title{NA61/SHINE results on Bose-Einstein correlations}

\ShortTitle{L\'evy HBT results at NA61/SHINE}

\author{\speaker{Barnab\'as P\'orfy for the NA61/SHINE collaboration}\\
        Wigner Research Centre for Physics, E\"otv\"os Lor\'and University, Budapest, Hungary\\
        E-mail: \email{bporfy@cern.ch}}

\abstract{One of the main goals of NA61/SHINE is the investigation of the phase diagram of strongly interacting matter. NA61/SHINE observes collisions of various nuclei at different energies, allowing to study the same phenomena and observables in vastly different conditions. One of the observables related to the quark-hadron transition is the Bose-Einstein momentum correlation function of identical pions, related to the space-time structure of pion emission. In this paper we report on such measurements in Be+Be collisions at an SPS beam momentum of 150\textit{A} GeV/\textit{c}. Our correlation functions can be statistically well described with L\'evy-distributed sources, hence we also study the $m_T$ dependence of the L\'evy source parameters, and discuss their possible interpretations.}

\FullConference{Corfu Summer Institute 2018 "School and Workshops on Elementary Particle Physics and Gravity"\\
		(CORFU2018)\\
		31 August - 28 September, 2018\\
		Corfu, Greece}

\begin{document}

\section{Introduction}
NA61/SHINE, located in the North Area of the CERN SPS, is a fixed target experiment with multiple Time Projection Chambers (TPCs) which are covering the full forward hemisphere. This results in an outstanding tracking down to $p_T = 0$ GeV/\textit{c}. Centrality of the collisions can be determined by the Projectile Spectator Detector (PSD), a modular calorimeter located on the beam axis, which is measuring the forward energy.

To study the phase diagram of the strongly interacting matter, one has to analyse collisions detected at different temperatures and baryon-chemical potentials. In this analysis we study Be+Be collisions at 150\textit{A} GeV/\textit{c}, which is a rather small system with low collision energy, hence different insights may be gained here than in large systems.  In our analysis to study the phase diagram of QCD we are measuring quantum-statistical correlations, in other words Bose-Einstein correlation functions, of identical pion pairs. This has a connection to the search for the critical point, as explained in the following.

The Critical End Point (CEP) of the QCD transition in the phase diagram may be characterized by its critical exponents. The critical exponent $\eta$ describes spatial correlations, which show power-law structure ($\sim r^{-(d-2+\eta)}$, where \textit{d} represents the number of spatial dimensions) at the critical point. With the assumption of QCD being in the universality class of the 3D Ising model~\cite{Halasz:1998qr,Stephanov:1998dy} one can obtain a conjectured value of $\eta$ at the CEP. From the 3D Ising model or from that with an external random field, the value of $\eta$ is at or below $0.5 \pm 0.05$~\cite{El-Showk2014,Rieger:PhysRevB.52.6659}. The common technique to extract the femtometer scale structures of the pion emitting source is to assume a Gaussian-type of source, but recent and some not-so-recent results~\cite{Csanad:2005nr,Adler:2006as,Adare:2017vig} show that alternative approaches should be considered. Even without a critical point, an expanding medium may lead to an increasing mean free path which in turn may lead to anomalous diffusion and L\'evy distributed sources~\cite{Adare:2017vig,Csanad:2007fr}, which represent a generalization of  Gaussian distributions. Symmetric L\'evy distributions are characterized by the L\'evy exponent $\alpha$ which determines the source shape, and a scale $R$ which relates to the homogeneity length of the expanding source. The symmetric L\'evy distribution is defined as follows:

\begin{equation}
\mathcal{L}(\alpha,R,r)=\frac{1}{(2\pi)^3} \int d^3q e^{iqr} e^{-\frac{1}{2}|qR|^{\alpha}}.
\end{equation}
In two special cases, the distribution can be analytically expressed: for $\alpha	= 1$ the result is a Cauchy distribution, for $\alpha = 2$ we get a Gaussian distribution. The L\'evy distribution is a suitable ``interpolator'' between these two, and even lower $\alpha$ values; but more importantly, there are physical processes that may indeed lead to these distributions~\cite{Csanad:2007fr,Csorgo:2004sr,Csorgo:2005it}. The main difference between a Gaussian distribution and a L\'evy with $\alpha < 2$ is the power-law tail$\sim r^{-(2-d+\alpha)}$ (where \textit{d} represents the number of spatial dimensions). Hence L\'evy distributions lead to spatial correlation functions with power-law tails similar to the ones present at the critical point, furthermore the critical spatial correlation exponent $\eta$ and the L\'evy exponent $\alpha$ are conjectured to be identical at the critical point~\cite{Csorgo:2005it}. Therefore a decrease in $\alpha$ in the vicinity of the critical point is expected, with $\alpha$ taking values at or below 0.5 at the CEP~\cite{Csorgo:2005it}. It is important to point out that anomalous diffusion and the vicinity of the CEP are possible alternative explanations for the appearance of L\'evy distributions.

The L\'evy exponent of the source distribution can be measured using Bose-Einstein correlation functions. These correlation functions become stretched type of exponentials for L\'evy sources:

\begin{equation}
C(q) = 1 + \lambda \cdot e^{-(qR)^\alpha}.\label{e:levyC0}
\end{equation}
One may note that if $\alpha = 1$ then the correlation function is exponential, while if $\alpha=2$, $C(q)$ becomes a Gaussian. The third physical parameter appearing in the above equation is the correlation strength $\lambda$ which is interpreted in the core-halo model. The correlation strength parameter $\lambda$ can then be expressed as

\begin{equation}\label{e:corr_str}
\lambda = (N_{\rm{core}}/(N_{\rm{core}} + N_{\rm{halo}})^2,
\end{equation}
where the source function is divided into two parts: the core consists of pions created close to the center (i.e. these are primordial pions or the decay products of extremely short lived resonances), while the halo contains the decay pions from long lived resonances. In the above equation, $N$ denotes their respective multiplicities in the given event class.

\section{Details of our analysis}
In this paper we discuss the measurements of one dimensional two-pion HBT correlation functions for identified pion pairs in Be+Be collisions at 150\textit{A} GeV/\textit{c} with 0-20\% centrality. We investigate positive pion pairs ($\pi^+\pi^+$),
negative pion pairs ($\pi^-\pi^-$), as well as their combined sample to increase statistics. Event and track quality cuts were applied before particle identification (PID). The PID was done by measuring the energy loss in the TPC gas ($dE/dx$) and comparison with the Bethe-Bloch curves. The correct centrality was selected by measuring the forward energy with the PSD. The measured pion pairs were grouped into four average transverse momentum bins ranging from 0 to 600 MeV/\textit{c}. All the mentioned settings were varied systematically, to obtain an a posteriori estimate of systematic uncertainties.

Due to the final state effect of the Coulomb-repulsion of like-sign charged particles, a correction has to be applied in the analysis. The general method for this is the application of a Coulomb correction. It is important to note that the Coulomb effect in this analysis is not very far from the Gamow factor due to the small system size, albeit a more elaborate treatment is still required. The calculation of the Coulomb correction for L\'evy sources may be obtained through a complicated numerical integral~\cite{Adare:2017vig}, but in our case, the correction is not expected to depend strongly on the values of the correlation function's shape parameter $\alpha$. Ergo one can use the approximative formula valid for $\alpha = 1$, published in Ref.~\cite{Sirunyan:2017ies}:

\begin{align}
K_{\textrm{Coulomb}}(q) &= \textrm{Gamow}(q)\cdot \left(1+\frac{\pi\eta q\frac{R}{\hbar c}}{1.26+q \frac{R}{\hbar c}}\right),\textnormal{ where}\label{e:coulcorr}\\
\textrm{Gamow}(q) &= \frac{2\pi\eta(q)}{e^{2\pi\eta(q)-1}} \;\textnormal{ and }\;
\eta(q) = \alpha_{\textrm{QED}}\cdot\frac{\pi}{q}.
\end{align}
Thus Eq.~\eqref{e:levyC0}, coupled with the Sinyukov-type of Coulomb-treatment~\cite{Sinyukov:1998fc}, modifies to the following expression:

\begin{equation}\label{e:fitfunc}
C(q) = N\cdot\left(1-\lambda + \left(1+e^{-(qR)^{\alpha}}\right)\cdot \lambda	\cdot K_{\textrm{Coulomb}}(q)\right)
\end{equation}
which is the fitting function we use in the analysis.

\section{Results and discussion}
The three physical parameters ($\alpha,\;\lambda\; \rm{and}\; R$) were measured in four bins of pair transverse momentum  $K_T$, to investigate their transverse momentum dependence. The three mentioned parameters were obtained through fitting the measured correlation functions with the formula of  Eq. \eqref{e:fitfunc}. In this proceedings contribution we report on the transverse mass dependence of $\alpha$, $\lambda$ and $R$, where transverse mass is expressed as $m_T=\sqrt{m^2+K_T^2}$, with $m$ being the pion mass.

\begin{figure}
     \centering
     \includegraphics[width=.9\textwidth]{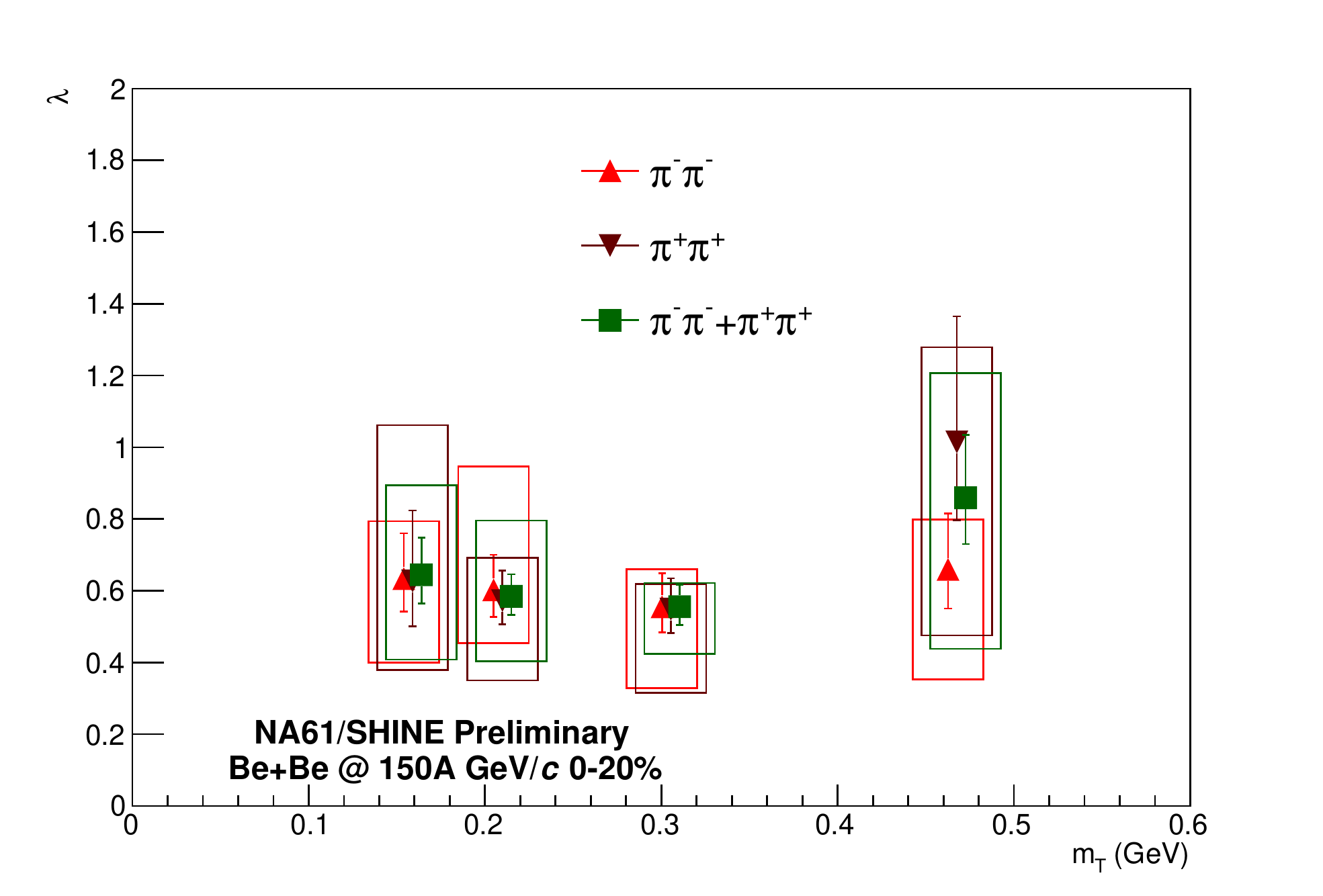}
     \caption{Fitting parameter $\lambda$ for three different pair samples, for 0-20\% centrality, as a function of $m_T$. For each bin, the results are slightly shifted horizontally for visibility. Boxes denote systematic, bars represent statistical uncertanities.}
     \label{fig:lambda}
\end{figure}

The first parameter we investigate is the correlation strength parameter $\lambda$, defined in Eq.~\eqref{e:corr_str}. The transverse mass dependence of $\lambda$ is shown in Fig.~\ref{fig:lambda}. One may observe that this parameter is slightly dependent on $m_T$, but mostly constant in the investigated range. When compared with measurements in RHIC Au+Au collisions~\cite{Adare:2017vig,Abelev:2009tp} and in SPS Pb+Pb interactions~\cite{Beker:1994qv,Alt:2007uj} an interesting phenomenon appears. In the case of SPS experiments, there are no visible ``holes'' at lower $m_T$ values, but  in the case of RHIC experiments, these appear. This ``hole'' was interpreted in Ref.~\cite{Adare:2017vig} to be a sign of in-medium mass modification. Our results, at the given statistical precision, do not indicate the presence of such a low-$m_T$ hole. Furthermore, it is important to note that our values for $\lambda$ are significantly smaller than unity, which might imply that a significant fraction of pions are the decay products of long lived resonances or weak decays. Finally, let us observe that all samples (positive pairs, negative pairs, combined sample) are consistent with each other.

\begin{figure}
     \centering
     \includegraphics[width=.9\textwidth]{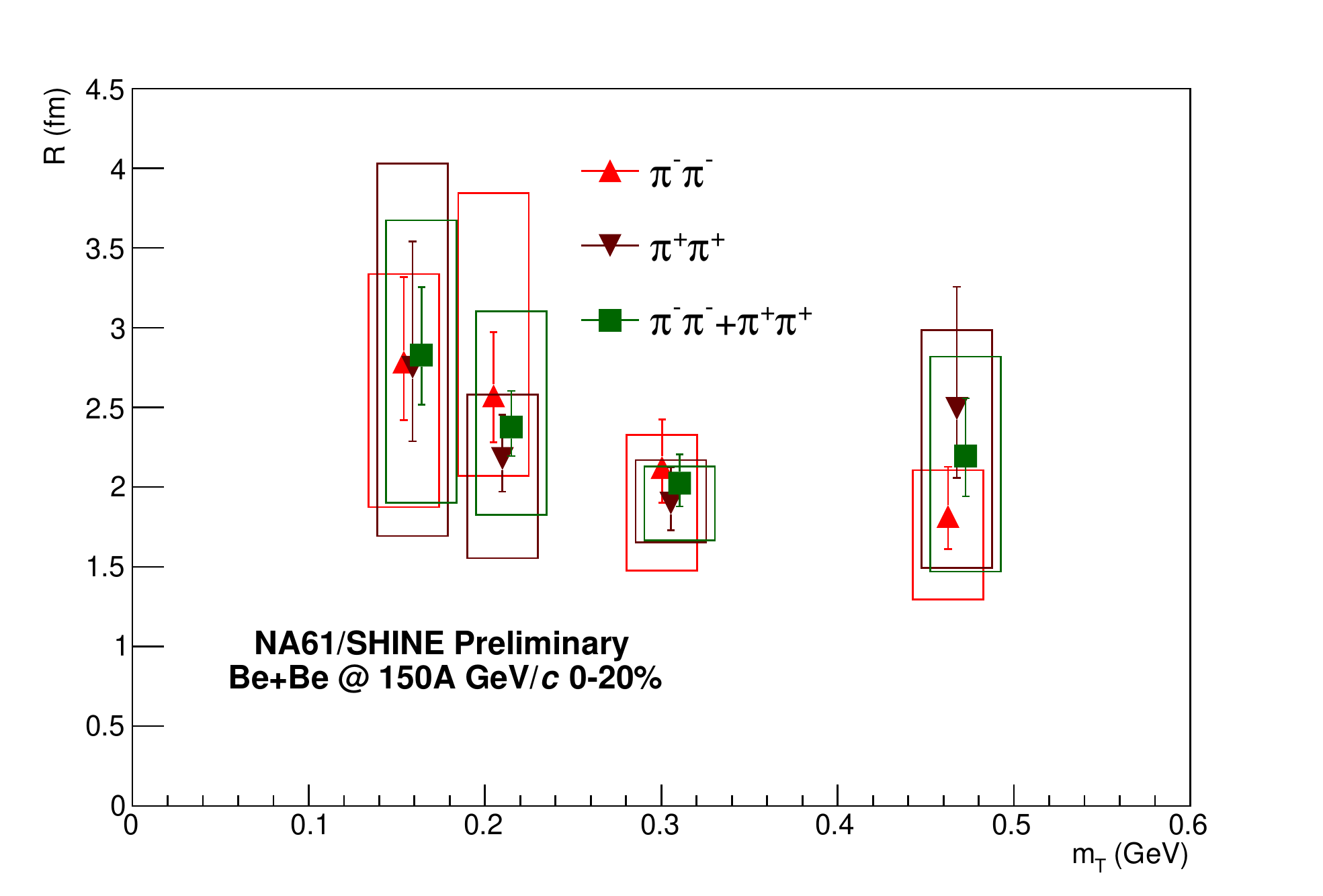}
     \caption{Fitting parameter $R$ for three different pair samples, for 0-20\% centrality, as a function of $m_T$. For each bin, the results are slightly shifted horizontally for visibility. Boxes denote systematic, bars represent statistical uncertanities.}
     \label{fig:R}
\end{figure}

The L\'evy scale parameter $R$ determines the length of homogeneity of the pion emitting source. From simple hydrodynamical models~\cite{Csorgo:1995bi,Csanad:2009wc} one obtains a $R \propto 1/\sqrt{m_T}$ type of transverse mass dependence. In Fig.~\ref{fig:R} a slight decrease for higher $m_T$ values may be observed. This might be the result of radial flow, as apparent in the mentioned hydrodynamical models. Let us furthermore mention, that the magnitude of our results is similar to homogeneity length values observed in $pp$ collisions at the LHC~\cite{Sirunyan:2017ies}.

\begin{figure}
     \centering
     \includegraphics[width=.9\textwidth]{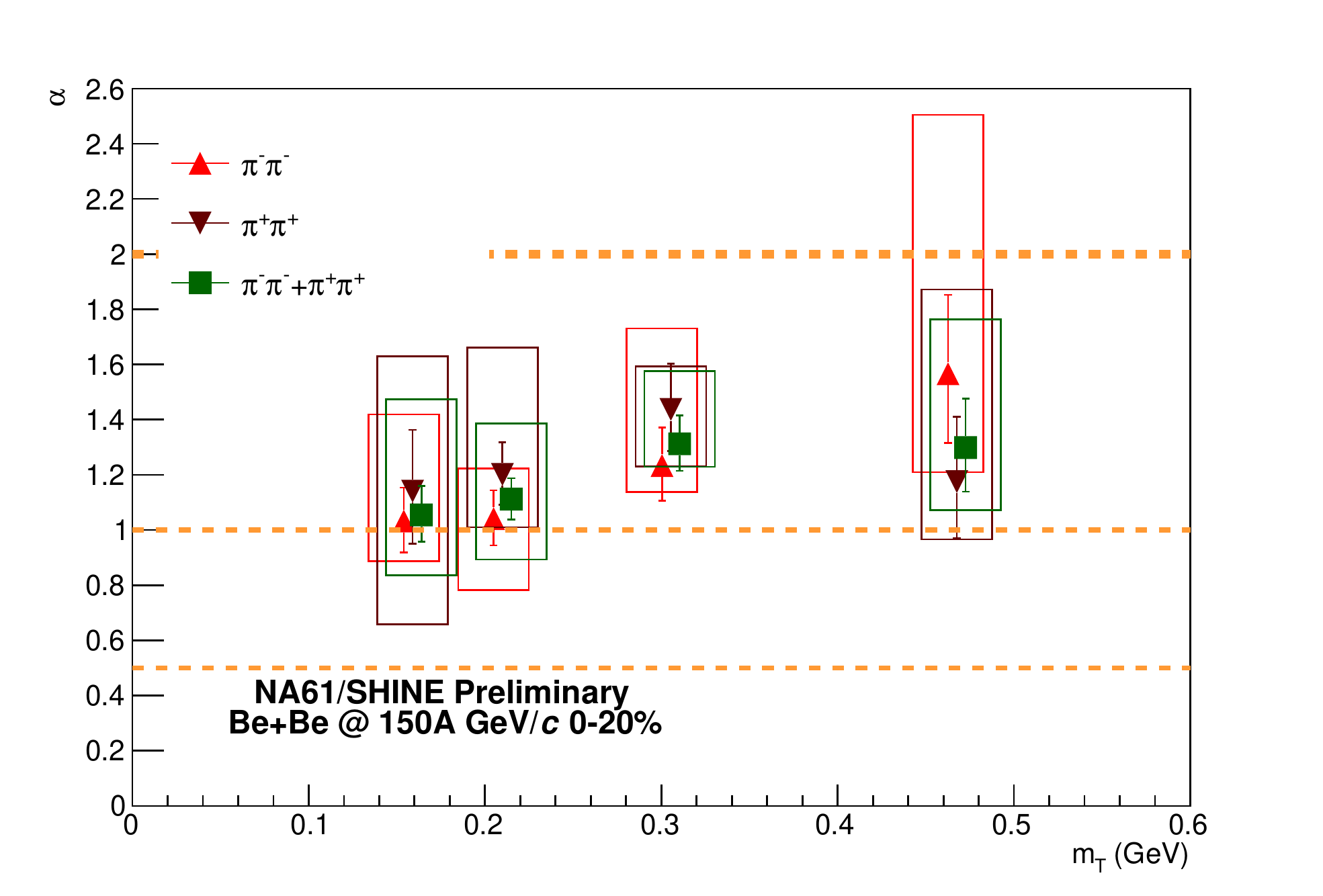}
     \caption{Fitting parameter $\alpha$ for three different pair samples, for 0-20\% centrality, as a function of $m_T$. For each bin, the results are slightly shifted horizontally for visibility. Boxes denote systematic, bars represent statistical uncertanities.}
     \label{fig:alpha}
\end{figure}

The L\'evy stability exponent $\alpha$ determines the shape of the source. Our results, indicated in Fig.~\ref{fig:alpha}, yield values for $\alpha$ scattering between 1.0 and 1.5, and are significantly lower than the Gaussian ($\alpha = 2$) case, and also significantly higher than the conjectured CEP value ($\alpha \leq 0.5$). Let us furthermore note that our values for $\alpha$ are similar to those obtained in Au+Au collisions at RHIC~\cite{Adare:2017vig}.

\section{Summary and outlook}

We measured one-dimensional two-pion Bose-Einstein correlation functions in 0-20\% centrality Be+Be collisions at 150\textit{A} GeV/\textit{c} at NA61/SHINE. We discussed the transverse mass dependences of the L\'evy source parameters. Results on the L\'evy scale parameter $\alpha$ showed a significant deviation from Gaussian sources, and are not in the vicinity of the conjectured value at the critical point. The L\'evy scale parameter $R$ shows a slight decrease with $m_T$, possibly explained by a non-negligible radial flow. The correlation strength parameter $\lambda$ does not show significant $m_T$ dependence, but indicates that a significant fraction of pions originate from resonance and weak decays. With these results at hand, we plan to measure Bose-Einstein correlations in larger systems as well as at smaller energies, to investigate the excitation function and system size dependence of the discussed physical parameters.

\section*{Acknowledgements}
The author would like to thank the NA61/SHINE collaboration, as well as NKFIH grant FK123842.

\end{document}